\documentclass[sigconf,nonacm]{acmart}

\usepackage{booktabs}
\usepackage{microtype}
\usepackage{xspace}
\usepackage{multirow}
\usepackage{balance}
\usepackage{enumitem}
\usepackage{tikz}

\newcommand{\aidev}{AIDev\xspace}
\newcommand{\ie}{\textit{i.e.}\xspace}
\newcommand{\eg}{\textit{e.g.}\xspace}
\newcommand{\etal}{\textit{et al.}\xspace}

\begin{document}

\title{Habituation at the Gate: Rising Approval and Declining Scrutiny in Human Review of AI Agent Code}

\author{Haoran Yu}
\affiliation{%
  \institution{Independent Researcher}
  \city{Seattle}
  \state{WA}
  \country{USA}}
\email{haoranyu889@gmail.com}

\author{Lifei Liu}
\affiliation{%
  \institution{Independent Researcher}
  \city{Seattle}
  \state{WA}
  \country{USA}}
\email{lliu.lifei@gmail.com}

\author{Xiaochong Jiang}
\affiliation{%
  \institution{Independent Researcher}
  \city{Seattle}
  \state{WA}
  \country{USA}}
\email{jiang.xiaoc@northeastern.edu}

\author{Yuwen Jia}
\affiliation{%
  \institution{Independent Researcher}
  \city{Santa Clara}
  \state{CA}
  \country{USA}}
\email{michelle2010tx@gmail.com}

\author{Su Wang}
\affiliation{%
  \institution{Carnegie Mellon University}
  \city{Pittsburgh}
  \state{PA}
  \country{USA}}
\email{suwang@alumni.cmu.edu}

\author{Pin Qian}
\affiliation{%
  \institution{Carnegie Mellon University}
  \city{Pittsburgh}
  \state{PA}
  \country{USA}}
\email{pqian@alumni.cmu.edu}

\author{Yihang Chen}
\affiliation{%
  \institution{Georgia Institute of Technology}
  \city{Atlanta}
  \state{GA}
  \country{USA}}
\email{ychen3726@gatech.edu}

\begin{abstract}
As AI coding agents (\eg GitHub Copilot, Devin, OpenAI Codex, Cursor) submit
pull requests to open-source repositories at scale, a key question
arises: do human reviewers gradually lower their scrutiny for AI-generated
code over time?
We conduct a longitudinal within-reviewer analysis using the \aidev dataset,
studying 400 repeat reviewers who collectively submitted 11,429 reviews
over a seven-month observation period.
Comparing each reviewer's early and late review episodes, we observe a
population-level shift in approval rate from \textbf{30.1\%} to
\textbf{36.8\%} (Wilcoxon signed-rank $p < 10^{-6}$ on paired shifts).
Pooled by within-reviewer experience decile, the cumulative gap reaches
\textbf{+14.5\,pp} from first to tenth decile.
This shift is experience-driven (persists after controlling for calendar
time), agent-specific (human PR approval rates decline over the same period),
and not explained by PR difficulty (median PR size is flat).
However, review latency \emph{increases} rather than decreases (+3.5$\times$),
while inline comment volume \emph{decreases} ($-$22\%, $p=0.0014$),
suggesting reviewers spend more time in queue but less time actively
inspecting code.
The combination of rising approval, declining comment effort, and
increasing queue time is most consistent with reflexive habituation
under growing workload rather than rational trust calibration alone.
\end{abstract}



\maketitle

\section{Introduction}

Code review is the primary human gate that separates an AI agent's output from
deployed software~\cite{bacchelli2013,rigby2013}.
This gate is only meaningful if reviewers maintain consistent scrutiny.
Prior work examines whether AI-generated PRs pass review~\cite{githubcopilot2023}
and has explored the inverse direction of using AI as a reviewer of human
code~\cite{li2022codereviewer,tang2024codeagent},
but almost no work asks whether human reviewers \emph{change their behavior}
as they accumulate experience with AI agents.

The \emph{rubber-stamping} hypothesis posits that reviewers, after repeatedly
approving agent-generated code with few defects, gradually reduce their
inspection effort and begin approving PRs reflexively.
This mirrors habituation effects documented in security auditing,
aviation checklists, and automated-testing interactions~\cite{cassee2020,wessel2020}.
If it occurs for AI agent PRs, it would undermine the human oversight layer
that AI-safety arguments often rely upon.

Our central question is:
\textbf{Do individual reviewers become systematically more approving of AI
agent pull requests as they accumulate reviewing experience?}

Concretely, we contribute:
\begin{itemize}[leftmargin=1.5em,topsep=2pt,itemsep=1pt]
  \item A within-reviewer longitudinal analysis of 11,429 reviews from
        400 repeat reviewers in the \aidev dataset~\cite{li2026aidev}.
  \item Evidence of a statistically significant population-level increase
        in approval rates (+14.5\,pp across experience deciles), driven by
        reviewer experience rather than calendar time, accompanied by a
        22\% decline in inline review comments ($\rho=-0.556$ with
        approval shift).
  \item A calendar-based human PR control showing that the trend is
        agent-specific: human PR approval rates decline over the same period.
\end{itemize}

We cannot establish causality: agent code may genuinely improve over the
observation window (all four agents received updates during 2025),
confounding reviewer behavior change with rational re-calibration.
However, the joint pattern of rising approval and declining comment effort
provides suggestive evidence toward habituation rather than pure rational
updating.

\section{Method}

\subsection{Dataset}

We use \aidev~\cite{li2026aidev}, a dataset of pull requests submitted by
AI coding agents to GitHub repositories with at least 100 stars.
The dataset contains \textbf{16,895 human reviews} across \textbf{2,494
unique reviewers}, covering five agent systems: GitHub Copilot Autofix,
Devin (Cognition AI), OpenAI Codex CLI, Cursor, and Claude Code
(Anthropic).
We focus our per-agent analysis on the four agents with sufficient
repeat-reviewer coverage; Claude Code contributes $<$2\% of
repeat-reviewer reviews.
For this study we focus on the \textbf{400 repeat reviewers} who each
reviewed 10 or more agent PRs, yielding \textbf{11,429 reviews}.
Within this cohort, \textbf{52 heavy reviewers} contributed 50 or more
reviews each, with individual observation spans up to 178 days
(the full dataset covers 207 days).

Each review record contains: reviewer identifier, agent identity, review
outcome (\textsc{approved}, \textsc{changes\_requested},
\textsc{commented}), review timestamp, and associated PR metadata.

\subsection{Longitudinal Design}

For each repeat reviewer we sort their reviews chronologically and split
them at the temporal midpoint into an \emph{early} and a \emph{late}
episode.
We compute per-reviewer approval rates ($\mathrm{AR}_{\mathrm{early}}$,
$\mathrm{AR}_{\mathrm{late}}$) and the individual shift
$\Delta\mathrm{AR} = \mathrm{AR}_{\mathrm{late}} - \mathrm{AR}_{\mathrm{early}}$.

For the decile analysis, we pool all reviews from repeat reviewers,
order them by the reviewer's \emph{within-reviewer review index} (1 =
first review of this reviewer, $n$ = latest), map indices to deciles,
and compute the mean approval rate per decile.

\subsection{Proxy Cross-Agent Control}

To probe whether any shift is reviewer-general or agent-specific, we
identify \textbf{108 reviewers} who reviewed PRs from at least two
distinct agents.
For each such reviewer we compute their approval rate shift separately
for each agent they reviewed.
If the shift were agent-specific (\eg because one agent's code genuinely
improved), we would expect the shift magnitude to differ markedly
across agents.

\subsection{Outcome Measures}

\begin{itemize}[leftmargin=1.5em,topsep=2pt,itemsep=1pt]
  \item \textbf{Approval rate (AR):} fraction of reviews with outcome
        \textsc{approved}.
  \item \textbf{Change-request rate (CRR):} fraction with outcome ``changes requested.''
  \item \textbf{Review latency:} hours between PR opening and review
        submission (median per episode).
\end{itemize}

\noindent
We use the Wilcoxon signed-rank test on paired
($\mathrm{AR}_{\mathrm{early}}$, $\mathrm{AR}_{\mathrm{late}}$) values
across all 400 reviewers.

\section{Results}

\subsection{Population-Level Longitudinal Trend}

\begin{table}[t]
\caption{Approval rate by experience decile (pooled repeat reviewers).
Decile 1 = reviewer's first reviews; Decile 10 = latest reviews.}
\label{tab:decile}
\small
\begin{tabular}{@{}lcc@{}}
\toprule
Decile & AR (\%) & CRR (\%) \\
\midrule
1 (earliest)  & 27.9 & 11.2 \\
2             & 27.1 & 11.3 \\
3             & 30.0 &  8.0 \\
4             & 28.1 &  7.9 \\
5             & 33.3 &  6.2 \\
6             & 35.6 &  7.0 \\
7             & 32.6 &  7.3 \\
8             & 34.2 &  6.6 \\
9             & 38.6 &  6.9 \\
10 (latest)   & 42.4 &  5.6 \\
\midrule
\multicolumn{2}{@{}l}{\textbf{Total shift}} &  \\
\multicolumn{3}{@{}l}{\textbf{+14.5\,pp} AR (27.9\%$\to$42.4\%); \textbf{$-$5.6\,pp} CRR} \\
\bottomrule
\end{tabular}
\end{table}

Table~\ref{tab:decile} shows an overall increase in approval rate and
decrease in change-request rate across experience deciles, with local
non-monotonicity due to binning noise.
The early-vs-late split analysis (400 reviewers, median 17 reviews each)
shows aggregate AR rising from \textbf{30.1\%} to \textbf{36.8\%}
(+6.7\,pp), with the change-request rate declining from 11.2\% to
5.6\%.\footnote{Aggregate rates are review-weighted (pooling all reviews).
Per-reviewer means give similar results: 30.5\% $\to$ 36.6\% (+6.1\,pp).}
The Wilcoxon signed-rank test on paired shifts yields
$p < 10^{-6}$, confirming the trend is not due to a small number of
influential reviewers and that the decile trend reflects genuine
within-reviewer shifts rather than reviewer-composition artifacts.
With $N=400$ pairs, the test has 80\% power to detect Cohen's $d \geq 0.14$
(approximately 2.1\,pp given observed $\mathrm{SD}(\Delta\mathrm{AR})=0.25$);
our observed $d=0.25$ is well above this threshold.

At the individual level, \textbf{52\% of reviewers} showed a positive
$\Delta\mathrm{AR}$ (\ie became more approving) while only 28\% became
less approving and 20\% showed no change ($\Delta\mathrm{AR} = 0$).

\newcommand{\arpt}[2]{\filldraw[black] (#1,{(#2-27)*1.333}) circle (1.6pt);}
\newcommand{\crpt}[2]{\filldraw[black!45] (#1,{(#2-5)*2.0+0.5}) rectangle +(2pt,2pt);}
\begin{figure}[t]
\centering
\begin{tikzpicture}[x=0.56cm, y=0.22cm]
  \foreach \pct/\ytk in {28/1.333, 30/4.0, 32/6.667, 35/10.667, 38/14.667, 42/20}{
    \draw[gray!25,very thin] (0,\ytk) -- (9,\ytk);
  }
  \draw[->,black!60] (-0.1,0) -- (9.5,0) node[right,font=\tiny]{decile};
  \draw[->,black!60] (0,-0.5) -- (0,21.5);
  \foreach \pct/\ytk in {28/1.333, 30/4.0, 35/10.667, 40/17.333}{
    \node[left,font=\tiny] at (-0.1,\ytk) {\pct\%};
  }
  \foreach \x in {0,1,...,9}{
    \node[below,font=\tiny] at (\x,-0.3) {\the\numexpr\x+1\relax};
  }
  \draw[black,thick] (0,{(27.9-27)*1.333}) -- (1,{(27.1-27)*1.333})
    -- (2,{(30.0-27)*1.333}) -- (3,{(28.1-27)*1.333}) -- (4,{(33.3-27)*1.333})
    -- (5,{(35.6-27)*1.333}) -- (6,{(32.6-27)*1.333}) -- (7,{(34.2-27)*1.333})
    -- (8,{(38.6-27)*1.333}) -- (9,{(42.4-27)*1.333});
  \arpt{0}{27.9}\arpt{1}{27.1}\arpt{2}{30.0}\arpt{3}{28.1}\arpt{4}{33.3}
  \arpt{5}{35.6}\arpt{6}{32.6}\arpt{7}{34.2}\arpt{8}{38.6}\arpt{9}{42.4}
  \draw[black!45,dashed,thick]
    (0,{(11.2-5)*2.0+0.5}) -- (1,{(11.3-5)*2.0+0.5})
    -- (2,{(8.0-5)*2.0+0.5}) -- (3,{(7.9-5)*2.0+0.5}) -- (4,{(6.2-5)*2.0+0.5})
    -- (5,{(7.0-5)*2.0+0.5}) -- (6,{(7.3-5)*2.0+0.5}) -- (7,{(6.6-5)*2.0+0.5})
    -- (8,{(6.9-5)*2.0+0.5}) -- (9,{(5.6-5)*2.0+0.5});
  \draw[black,thick] (5.5,19.5) -- (6.5,19.5);
  \filldraw[black] (6.0,19.5) circle (1.5pt);
  \node[right,font=\tiny] at (6.6,19.5) {Approval rate};
  \draw[black!45,dashed,thick] (5.5,17.5) -- (6.5,17.5);
  \node[right,font=\tiny] at (6.6,17.5) {CR rate (rescaled)};
\end{tikzpicture}
\caption{Approval rate (solid circles, left axis) and change-request rate
(dashed, rescaled $2\times$ for visualization) across 10 experience deciles.
Approval rises overall from 27.9\% to 42.4\% (+14.5\,pp) with local
fluctuations; CR rate declines from 11.2\% to 5.6\%.}
\label{fig:decile}
\end{figure}
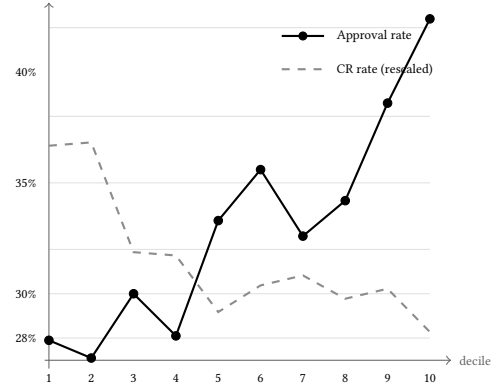

Review latency also changes markedly: reviewers take a median of
\textbf{3.9 hours} in their early episode but \textbf{13.5 hours} in
their late episode.
This apparently contradictory finding---approving more yet taking longer---may
reflect reviewers accumulating a larger review backlog over time,
deprioritizing routine agent PRs, or changes in PR submission timing
rather than slower active inspection.

\subsection{Review Effort: Inline Comment Analysis}

To probe whether rising approval reflects declining inspection effort,
we analyze inline review comments from the same reviewers.
Pooled by experience decile, the mean number of inline comments per
review drops from \textbf{1.01} (first decile) to \textbf{0.79} (tenth
decile), a 22\% decline.
Total comment word count per review decreases from 18.6 to 13.5 words
($-$28\%).
Both declines are statistically significant in paired early-vs-late
tests (Wilcoxon $p=0.0014$ for comment count, $p=0.0029$ for word
count).

The within-reviewer correlation between approval-rate shift and
comment-effort shift is strongly negative: Spearman
$\rho=-0.556$ ($p < 10^{-4}$) for $\Delta\mathrm{AR}$ vs.\
$\Delta(\text{comments/review})$.
Among reviewers who drifted upward ($\Delta\mathrm{AR} > +0.10$),
inline comments decreased by 0.48 per review on average; among stable
reviewers ($|\Delta\mathrm{AR}| < 0.05$), comments were unchanged
($+0.10$).
This pattern---more approvals accompanied by less commenting---is
consistent with declining inspection effort rather than
unchanged effort with a rationally raised threshold.

\subsection{Per-Agent Breakdown}

\begin{table}[t]
\caption{Early-to-late approval rate shift per agent (repeat reviewers
only). $n$ = number of repeat reviewers with $\geq$5 reviews for that
agent.}
\label{tab:agent}
\small
\begin{tabular}{@{}lcrrr@{}}
\toprule
Agent        & $n$   & AR$_\text{early}$ (\%) & $\Delta$AR (pp) & 95\% CI \\
\midrule
Copilot  & 241 & 28.1 & +9.1 & [+5.8, +12.3] \\
Devin    &  95 & 33.0 & +3.8 & [$-$0.8, +8.3] \\
Codex    &  51 & 35.5 & +1.8 & [$-$6.4, +9.6] \\
Cursor   &  18 & 39.8 & $-$1.4 & [$-$14.8, +10.3] \\
\bottomrule
\end{tabular}
\end{table}

Table~\ref{tab:agent} reports the approval-rate shift for each agent.
Copilot reviewers exhibit the strongest positive shift (+9.1\,pp),
consistent with Copilot being the most mature and widely deployed agent
in our observation window.
Devin shows a moderate shift (+3.8\,pp).
Codex and Cursor are near-stable ($\leq$2\,pp change); however, both have
small repeat-reviewer populations ($n \leq 51$), limiting statistical power.
We caution against reading these between-agent differences causally:
Copilot's larger shift may reflect the larger and more diverse pool of
Copilot reviewers rather than a property of the agent itself.

\subsection{Proxy Cross-Agent Control}

To assess whether the trend is reviewer-specific or agent-specific,
we examine the 108 reviewers who reviewed PRs from two or more agents.
Among the 75 (reviewer, agent) pairs that each had at least five reviews,
the mean approval-rate shifts are:
Codex +9.4\,pp, Copilot +9.8\,pp, Devin +8.7\,pp.
Cursor ($n=10$ pairs) shows $-$10.3\,pp, but this estimate is
extremely noisy given the small sample.

The \emph{similarity} of shifts across agents for the same set of
reviewers is \emph{consistent with} a \textbf{reviewer-general}
phenomenon rather than one driven by a single agent's code quality
improving.
That is, a reviewer who becomes more approving of Copilot PRs also tends
to become more approving of Devin and Codex PRs.

We stress, however, that this cross-agent control is underpowered:
only 1--2 reviewer--agent pairs overlap for direct within-reviewer,
within-agent comparison, making a fully controlled test infeasible
with the current dataset.

\subsection{Calendar-Based Human PR Control}

To distinguish reviewer-general leniency from agent-specific adaptation,
we crawled review records for 6,618 human-authored PRs from the same
repositories via the GitHub API, obtaining 11,415 reviews from 1,851
reviewers.
Of these, \textbf{728 reviewers} appear in both the agent and human PR
review pools.

\begin{figure}[t]
\centering
\includegraphics[width=\columnwidth]{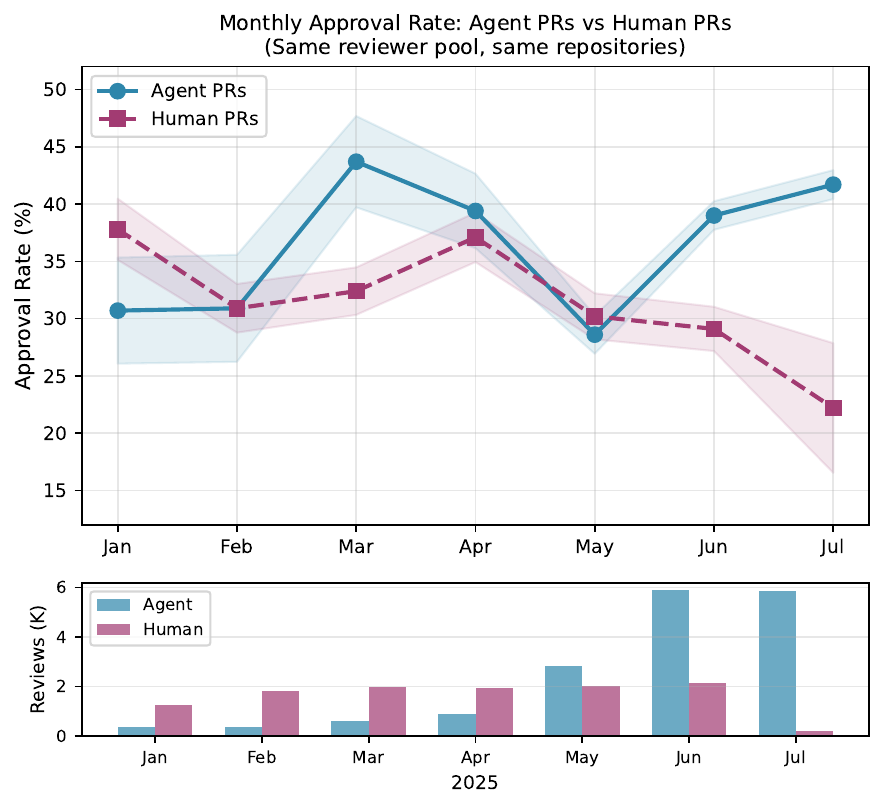}
\caption{Monthly approval rate for agent PRs (solid) vs.\ human PRs
(dashed) in overlapping repositories with 95\% binomial CI bands (shaded).
Same reviewer pool. Bottom panel shows review volume.
Human PR volume drops sharply in July ($n=207$); the late-period
divergence has wide uncertainty.}
\label{fig:calendar}
\end{figure}

Figure~\ref{fig:calendar} compares monthly approval rates for agent and
human PRs across the same reviewer population and repository set.
In January 2025, agent PRs were approved \emph{less} often than human PRs
(30.7\% vs.\ 37.8\%, $\Delta=-7.1$\,pp).
Over the following months, the two trends diverge: agent AR rises to
41.7\% by July while human AR declines to 29.1\% by June (22.2\% in July,
though with only $n=207$ reviews).
By June 2025, reviewers approve agent PRs 10\,pp more readily than human
PRs from the same repositories.

This calendar-based comparison strengthens the agent-specific
interpretation: if reviewers were simply becoming more lenient overall,
human PR approval rates should rise in tandem.
Instead, human AR is flat or declining while agent AR rises, consistent
with reviewers developing differential trust in agent-generated code.

We note two caveats: (1)~human PR review volume drops sharply after June
2025 ($n=207$ in July), making the late-period human AR estimate noisy;
and (2)~this is a population-level comparison, not a within-reviewer
paired test (the per-reviewer calendar split yields only 26 agent-side
reviewers with sufficient data in both periods).

\section{Discussion}

\subsection{What the Data Show}

We observe a consistent pattern: reviewers who accumulate experience
reviewing agent PRs become more approving over time (+14.5\,pp from first
to tenth decile).
Three sensitivity checks help locate the source of this shift:

\emph{Experience vs.\ calendar time.}
We fit a logistic regression separately for each of the 343 reviewers
with $\geq$10 reviews for whom the model converges (57 reviewers with
no outcome variation---52 never approved, 5 always approved---are
excluded), predicting approval from both within-reviewer experience
index and calendar date (days since January~1).
Pooling coefficients: experience carries the signal (mean coefficient
+0.11, 65\% of reviewers positive) while calendar time contributes
little (mean $-$0.007, 46\% positive).
The trend is driven by individual reviewer accumulation, not by a
global temporal shift.

\emph{PR size is flat.}
Monthly median PR size (lines changed) does not decline over the
observation window (Spearman $\rho=+0.02$, $p=0.009$, negligible magnitude).
Later agent PRs are not systematically simpler; the approval increase
cannot be explained by declining PR difficulty.

\emph{Agent-specific, not global leniency.}
The calendar-based human PR control (Figure~\ref{fig:calendar}) shows agent
AR rising while human AR declines over the same period.
If reviewers were simply becoming more lenient across all PRs, human PR AR
should rise in parallel. It does not.

\subsection{Three Interpretations}

These results are consistent with three accounts:
\textbf{(1)~Progressive trust calibration}---reviewers rationally raise
their approval threshold based on accumulated positive evidence;
\textbf{(2)~Deliberation under workload}---growing backlog pressure
produces longer latency but ultimately resolves toward approval
($\bar{\rho}=+0.08$ between latency and approval, $p \approx 0.21$);
\textbf{(3)~Reflexive habituation}---reviewers reduce inspection effort
after repeated positive experiences.

The inline comment analysis provides partial disambiguation.
The strong negative correlation between approval shift and comment-effort
decline ($\rho=-0.556$, $p < 10^{-4}$) is more consistent with~(3)
than~(1): under pure trust calibration, reviewers would approve more but
maintain inspection depth.
Instead, rising approval co-occurs with 28\% less commenting.
The latency increase is best explained by longer queue time but shorter
active review---the nuanced variant of~(3).
We cannot fully exclude~(1), as some effort reduction may be rational,
but the magnitude suggests effort reduction beyond rational updating.

\subsection{Per-Agent Asymmetry}

Among agents with $n>90$ reviewers, both show clear positive shift:
Copilot (+9.1\,pp, $n=241$) and Devin (+3.8\,pp, $n=95$).
Codex ($n=51$) and Cursor ($n=18$) are underpowered for reliable
agent-specific conclusions.

\subsection{Implications for Practice}

While the comment-effort analysis favors habituation over pure trust
calibration, the practical outcome is the same regardless of mechanism:
over time, a growing majority of reviewers approve agent PRs at higher
rates with less scrutiny.
Among those with meaningful change ($|\Delta\mathrm{AR}| \geq 0.05$),
65\% moved toward more approval and 35\% toward less.
Teams relying on human review as a quality gate should consider:
\begin{itemize}[leftmargin=1.5em,topsep=2pt,itemsep=1pt]
  \item \textbf{Rotation policies} that prevent any single reviewer from
        accumulating an excessive share of agent PR reviews.
  \item \textbf{Streak audits} that flag long consecutive approval runs
        from a single reviewer for secondary inspection.
  \item \textbf{Trend dashboards} that show reviewers their personal
        approval-rate trajectory alongside downstream defect data.
\end{itemize}

\subsection{Limitations}

The \aidev dataset covers only repositories with $>$100 stars;
enterprise repositories may differ.
Our observation window spans at most 207 days.
Cursor ($n=18$) and Codex ($n=51$) have small reviewer populations.
The human PR control (Figure~\ref{fig:calendar}) loses volume after
June 2025 ($n=207$ in July), adding noise to the late-period comparison.
We cannot rule out that the latency increase reflects systematic
changes in PR submission timing rather than reviewer behavior.
Crucially, we lack a direct measure of agent code quality over time;
if agents genuinely improved, rising approval would be rational rather
than habituative.

\section{Related Work}

Code review quality and efficiency have been studied extensively~\cite{bacchelli2013,rigby2013,mcintosh2016}.
Kalliamvakou \etal~\cite{kalliamvakou2014} warned of biases in mining
GitHub data, a caveat we inherit.
Wessel \etal~\cite{wessel2020} showed that code review bots affect
contributor behavior in OSS projects; our work extends this to AI coding
agents and reviewer-side behavior.
Kononenko \etal~\cite{kononenko2015} linked review quality and
participation to defect outcomes; tracking review quality longitudinally
would complement our outcome-based analysis.
Cassee \etal~\cite{cassee2020} found that CI adoption changes reviewer
behavior; AI agents represent a larger-scale intervention in the same vein.

\section{Conclusion}

We report a longitudinal within-reviewer analysis of human code
review behavior for AI agent pull requests.
Across 400 repeat reviewers and 11,429 reviews, we observe a statistically
significant increase in approval rates (+14.5\,pp across experience
deciles) and a corresponding decline in change-request rates.
This shift is experience-driven (not calendar-driven), agent-specific
(human PR approval rates decline over the same period in the same
repositories), and not explained by declining PR difficulty (median PR size
is flat).
The concurrent increase in review latency ($+3.5\times$) coupled with a
decline in inline comment effort ($-$22\%, $\rho=-0.556$ with approval
shift) suggests that reviewers spend more time in queue but less time
actively inspecting, consistent with reflexive habituation under growing
workload.
This pattern raises actionable concerns for teams relying on human
oversight of agentic software development.
We release our analysis scripts to support replication and extension.\footnote{\url{https://anonymous.4open.science/r/reviewer-behavioral-shift-FE86}}

\balance
\bibliographystyle{ACM-Reference-Format}

\end{document}